\documentclass[aps,prl,twocolumn,superscriptaddress,floatfix]{revtex4-1}
\usepackage{amsmath,amssymb,amsfonts,braket,graphicx,xcolor,mathtools,bm,xfrac,footmisc,paralist,mwe,siunitx}
\usepackage[colorlinks=true,allcolors=blue]{hyperref}
\usepackage[version=4]{mhchem}
\usepackage[T1]{fontenc} % encoding LY1 does not support \dj
\usepackage[varg]{txfonts}

\newcommand{\mi}{\mathrm{i}}
\newcommand{\me}{\mathrm{e}}

\newcommand{\abs}[1]{|#1|}

\def\equationautorefname~#1\null{Eq.~(#1)\null}
\allowdisplaybreaks

\begin{document}
%	\title{Spin-Selective Andreev Reflection in a Realistic Topological Nodal-Point Superconductor}
%	\title{Detection of Majorana Modes by Spin-Selective Andreev Reflection in a Multiply Edged Topological Nodal-Point Superconductor}
%	\title{Detection of Majorana Edge Modes by Spin-Selective Andreev Reflection without Overwhelming Spin-Triplet Pairing Correlations}
%	\title{Detection of Majorana Edge Modes without Overwhelming Spin-Triplet Pairing Correlations in Topological Nodal-Point Superconducting Systems}
	%\title{Detection of Majorana Edge Modes without Overwhelming Spin-Triplet Pairing Correlations from Topological Superconductivity in Antiferromagnetism}
%	\title{Finite Spin-Singlet Pairing Correlation of Majorana Edge Modes and Its Detection}
	\title{Spin Signature of Majorana Fermions in Topological Nodal-Point Superconductors}
	\author{Junjie Zeng}
	\affiliation{Institute for Structure and Function \& Department of Physics \& Chongqing Key Laboratory for Strongly Coupled Physics, Chongqing University, Chongqing 400044, P. R. China}
	
	\author{James Jun He}
	\affiliation{International Center for Quantum Design of Functional Materials (ICQD), Hefei National Research Center for Interdisciplinary Sciences at the Microscale, University of Science and Technology of China, Hefei, Anhui 230026, China}
	\affiliation{Hefei National Laboratory, Hefei, Anhui 230088, China}

    \author{Zhen Ning}
	\affiliation{Institute for Structure and Function \& Department of Physics \& Chongqing Key Laboratory for Strongly Coupled Physics, Chongqing University, Chongqing 400044, P. R. China}
	
	\author{Dong-Hui Xu}
	\email{donghuixu@cqu.edu.cn}
	\affiliation{Institute for Structure and Function \& Department of Physics \& Chongqing Key Laboratory for Strongly Coupled Physics, Chongqing University, Chongqing 400044, P. R. China}
    \affiliation{Center of Quantum materials and devices, Chongqing University, Chongqing 400044, P. R. China}
	
	\author{Rui Wang}
	\email{rcwang@cqu.edu.cn}
	\affiliation{Institute for Structure and Function \& Department of Physics \& Chongqing Key Laboratory for Strongly Coupled Physics, Chongqing University, Chongqing 400044, P. R. China}
	%\affiliation{Center for Computational Science and Engineering, Southern University of Science and Technology, Shenzhen 518055, P. R. China}
	\affiliation{Center of Quantum materials and devices, Chongqing University, Chongqing 400044, P. R. China}
	
	\date{\today}
	
	\begin{abstract}
In two-dimensional topological nodal superconductors, Majorana edge states have been conventionally believed to exhibit only spin-triplet pairing correlations. However, we reveal a substantial spin-singlet pairing component in Majorana edge states of antiferromagnetic topological nodal-point superconductors. This unexpected phenomenon emerges from the interplay between antiferromagnetic order and symmetry, resulting in Majorana edge states with a nearly flat band dispersion, deviating from the strictly flat band. Crucially, this phenomenon is detectable through spin-selective Andreev reflection, where the zero-bias conductance peaks are maximized when the spin of incident electrons is nearly antiparallel to that of Majorana edge excitations. This discovery unveils a unique spin signature for Andreev reflection resonances, advancing our fundamental understanding of spin-dependent mechanisms in topological superconductivity and representing a significant step towards the experimental detection of Majorana fermions.
%that Majorana edge states with broken chiral symmetry but nearly flat bands can exhibit a surprisingly strong spin-singlet pairing component. We reveal that this phenomenon arises due to  and can be detected through spin-selective Andreev reflection in topological nodal-point superconductors with antiferromagnetic order. We demonstrate that the zero-bias conductance peaks are maximized when the spin polarization of the incident electrons is nearly antiparallel to that of the Majorana edge excitations, revealing a unique spin signature for Andreev reflection resonances.
%In two-dimensional topological nodal superconductors, Majorana edge states with nearly flat bands exhibit a surprisingly strong spin-singlet pairing component compared with strictly flat band states. This phenomenon, which was previously overlooked, arises due to broken chiral symmetry and can be detected through spin-selective Andreev reflection in recently realized topological nodal-point superconductors with antiferromagnetic order. We demonstrate that the zero-bias conductance peaks are maximized when the spin polarization of the incident electrons is nearly antiparallel to that of the Majorana edge excitations, revealing a unique spin signature for Andreev reflection resonances. This finding advances our understanding of spin-dependent mechanisms in topological nodal superconductivity and represents a significant step towards the experimental detection of Majorana fermions.
	\end{abstract}
	
	\maketitle
	
	\textit{Introduction}.---In recent years, topological superconductors have attracted significant attention as a canonical example of topological phases of matter which host exotic Majorana quasiparticles. Majorana zero modes~(MZMs), exhibiting non-Abelian statistics, are predicted to exist in these systems--either as end states in one-dimensional~(1D) topological superconductors or as vortex bound states in two-dimensional~(2D) topological superconductors~\cite{Kitaev2001PU,Read:00,Volovik:99,Ivanov:01}. In addition to the fundamental physical interest, MZMs provide a practical avenue for information processing in future fault-tolerant quantum computers~\cite{Kitaev2003AP,Nayak2008RMP} as well. Topological superconductors also exhibit Majorana modes bound to their 1D or 2D boundaries, enforced by the bulk-boundary correspondence. Moreover, beyond the gapped topological superconductors, gapless nodal superconductors with nontrivial topology are another class of topological phases accommodating Majorana boundary excitations which terminate at the projection of bulk nodes~\cite{schnyder2015topological}. Topological nodal superconductors have been proposed in various superconducting materials with the need for unconventional Cooper pairing~\cite{Sato2006,Yang:14,Beri2010,Nayak2021NP}. Meanwhile, artificial topological nodal superconductivity is predicted to be realized in hybrid systems with conventional $ s $-wave pairing symmetry~\cite{Sato2010,Wong2013PRB}. A recent focus has centered on the realization of topological nodal superconductivity in hybrid systems by proximity-coupling antiferromagnetic (AFM) materials to $ s $-wave superconductors~\cite{Ladoprl2018,Brezicki2018,Zhang2019PRL}. Notably, antiferromagnets, devoid of a net magnetic moment, avoid significant exchange fields that could disrupt Cooper pairs. Very recently, 2D topological nodal-point superconductivity is experimentally reported in the AFM magnet/superconductor hybrid of Mn/Nb(110) by measuring low-temperature tunneling spectroscopy~\cite{Bazarnik2023NC}.

    %No material in nature exhibits topological superconductivity.
    %The basic characteristic of those Majorana fermions is to possess a large value of density of states in the vicinity of zero energy. Usually, the theoretical considerations are accompanied with chiral symmetry \cite{Wong2013PRB,He2018CP,Zhang2019PRL}. By this symmetry, the Majorana edge states on the boundary of a two-dimensional (2D) TNPSC are guaranteed to have flat bands. Consequently, this situation can be deemed as a trivial dimension generalization of the 1D nanowire TSC cases with in-gap zero-energy Majonara end states, if one loosely treats the transverse lattice momentum as a system parameter. However, for realistic material systems, the chiral symmetry is too strong to conserve because of the inevitable boundaries, especially as observed experimentally recently in the \ce{TaS2} system \cite{Nayak2021NP} and the \ce{Mn/Nb} system \cite{LoConte2022PRB,Bazarnik2023NC}.Nevertheless, before substantially harnessing it, its existence and basic properties might be the key issues demanding confirmation and clarification.
	Whether Majorana quasiparticles have been conclusively detected is still an open and controversial question in experimental investigations of topological superconductors. One of the most prominent features pertained to MZMs probably is the zero-bias conductance peaks (ZBCPs), from experiments of either scanning tunneling microscope measurements or transport processes in tunneling junctions. Therefore, many research works treat ZBCPs as the suggesting evidence for the occurrence of MZMs~\cite{mourik2012signatures,Das2012NP,nadj2014observation,Xu2015PRL,menard2017two,Chen2017SA,Wang2018S,kezilebieke2020topological,Vaitiekenas2020NP,manna2020signature,Liang2021PRX}. But whether a feature as ZBCPs adequately supports the existence of MZMs has not yet gained its full certainty, because the appearance of ZBCPs, can have other provenance, e.g., topologically trivial bound states \cite{Liu2012PRL,Pikulin2012NJP,Moore2018PRB,Chen2019PRL,Vuik2019SP,Juenger2020PRL,Prada2020NRP,Pan2020PRR,Sarma2021PRB}. % Additionally, the situations have been further complicated by the fact that the behaviors of the ZBCPs stemming from Majorana bound states depend on the dimensions and symmetries of the systems under study \cite{Tanaka2009PRL,Tanaka2009PRB,Yada2014JPSJ,Liu2015PRB}.
    % a general kind of 2D system with intrinsic antiferromagnetism and conventional superconducting pairing has been experimentally synthesized , which, confirmed with topological nontriviality, although in the absence of a bulk band gap, still is observed that edges of different kinds host Majorana modes with dispersions, esp., the zigzag edge can accommodate Majorana edge modes with nearly nondispersive bands. Beyond our original expectation, we find that the spin-singlet pairing correlation manifests itself with fairly conspicuous existence in both the zigzag and the ferromagnetic Majorana edge states. As a result, the usual conception would not happen that the ZBCPs in Andreev reflection become the most obvious when the spin polarizations of the inscattering electron and the Majorana mode are (about to) parallel, and is just replaced by its opposite scenario.

	On the other hand, MZMs leave unique fingerprints in spin-sensitive measurements~\cite{sticlet2012prl,He2014PRL,Haim2015prl,Sun2016PRL,Hu2016PRB,Jeon2017science,wang2021prl,jack2021detecting}, thus providing the means to distinguish them from trivial zero-energy states. The basic standing ground is that MZMs have spin-triplet pairing correlations~\cite{Liu2015PRB,Zhang2017PRB}. These pairing correlations can manifest in transport behaviors when the quasiparticles participate in Andreev reflection processes. Remarkably, MZM-induced spin-selective Andreev reflection was experimentally evidenced in detecting vortex core states of topological superconductivity in the topological insulator/superconductor heterostructures~\cite{Sun2016PRL}. In 2D topological nodal superconductors, the Majorana edge modes with strictly flat bands are believed to exhibit spin-triplet pairing correlations due to numerous MZMs residing on the sample boundaries~\cite{Yuan2017PRB,he2018magnetic}. Therefore, the spin-selective Andreev reflection induced by MZMs is also expected to appear in topological nodal superconductors. Considering that the topological nodal superconductivity with the AFM order is protected by the spatial-dependent symmetry, the spectrum features of Majorana edge modes would strongly depend on the edge geometry.

A natural question arises whether the interplay of AFM order and symmetry in topological nodal superconductors leads to exotic spin-resolved transport phenomena. In this Letter, we address this issue and uncover a previously overlooked scenario. We show that spin-singlet pairing correlations play a significant role due to Majorana edge modes manifesting in nearly flat bands instead of strictly flat bands.
%in Majorana edge modes connecting the projections of two bulk nodes, spin-triplet pairing correlations may not always dominate, allowing observable effects from spin-singlet pairing.
%We find that edge tunneling favors electrons with spin polarization antiparallel to that of Majorana edge modes.
Remarkably, the ZBCPs  in Andreev reflection are now strongest when the spin polarization of incident electrons is antiparallel to that of Majorana edge modes. This challenges the widely-held belief that ZBCPs are most prominent when the spin polarizations of incident electrons and Majorana edge modes are (nearly) parallel. Our findings not only unveil a novel spin signature in topological superconductivity, but also further enhance the understanding of the spin-dependent mechanism of Majorana fermions with further theoretical and experimental implications. %highlighting the risk of misinterpreting ZBCP behavior without considering this spin-dependent mechanism.

\begin{figure}[htp!]
		\centering
		\includegraphics[width=0.4\textwidth]{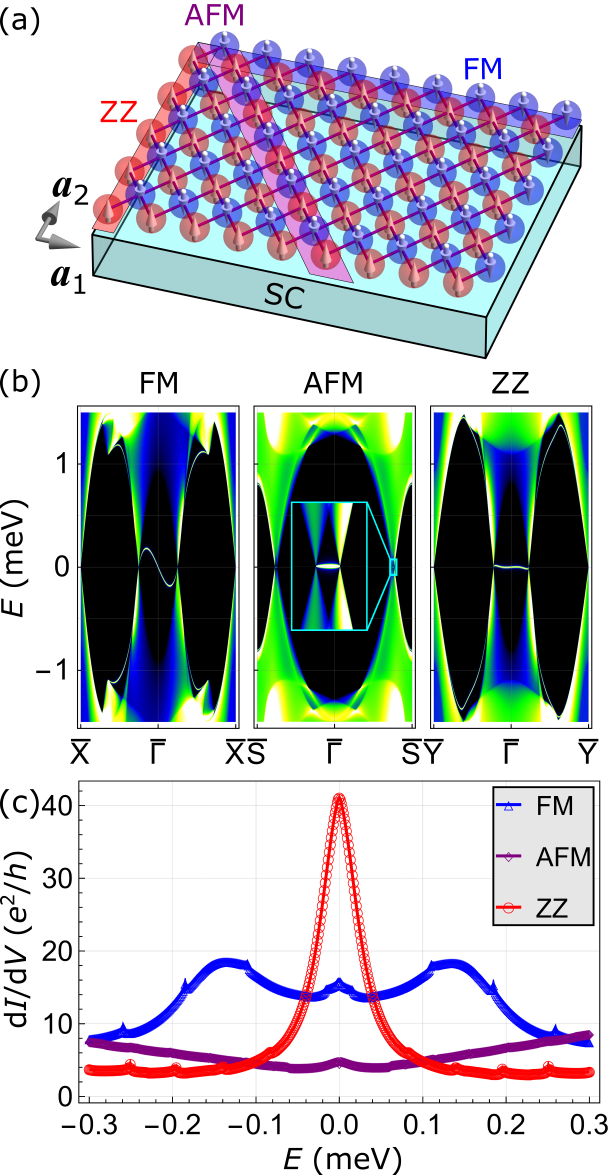}
		\caption{\label{fig1:basic_structure_and_behavior} Scenario of existence of the Majorana edge modes and their corresponding Andreev reflection behaviors. (a) Lattice structure of an AFM monolayer system on top of a conventional superconducting substrate, with lattice constants along the two orthogonal directions $ \abs{\bm a_2} = \sqrt{2} \abs{\bm a_1} = \sqrt{2} $, if we chose $ \abs{\bm a_1} = 1 $; the red and blue balls show the two sets of sublattices with opposite magnetic polarizations and connected by purple sticks are the 1st nearest neighbors; the three types of edges are labeled, respectively, by light blue, purple, and red background bands. (b) The existence of Majorana edge modes for each type of edge by the spectral function. (c) The ZZ edge mode gives rise to the most conspicuous peak signal in the process of Andreev reflection of a single-terminal tunneling junction.}
	\end{figure}

	\textit{Model and Basic Behavior in Andreev Reflection}.---The topological nodal-point superconductivity coexisting with antiferromagnetism in a 2D rectangular bisublatticed system [as shown in \autoref{fig1:basic_structure_and_behavior}(a)] is captured by the model \cite{Li2016NC,Rachel2017PRB,Bazarnik2023NC}
	\begin{align}\label{eq:Hamiltonian}
		H &= H_0 + H_\text{RSOC} + H_\text{AFM} + H_\text{SC}, \\
		H_0 &= \left(t \sum_{\braket{ij}} + t' \sum_{\braket{\braket{ij}}} + t'' \sum_{\braket{\braket{\braket{ij}}}}\right) c_i^\dagger \cdot c_j - \mu \sum_i c_i^\dagger \cdot c_i, \notag\\
		H_\text{RSOC} &= \mi t_\text{R} \sum_{\braket{ij}, \braket{\braket{ij}}} c_i^\dagger \cdot (\bm{s} \times \hat{\bm{n}}_{ij})_z \cdot c_j, \notag\\
		H_\text{AFM} &= J \sum_i c_i^\dagger \cdot (\me^{\mi \bm{x}_i\cdot \bm Q + \mi \pi/2} s_3) \cdot c_i, \notag\\
		H_\text{SC} &= \Delta \sum_i c_{i\uparrow}^\dagger c_{i\downarrow}^\dagger + \text{H.c.}, \notag
	\end{align}
	where the first term $ H_0 $ contains the chemical potential $ \mu $ and hoppings up to the 3rd nearest neighbors; the second term $ H_\text{RSOC} $ is the Rashba spin-orbit coupling (RSOC) measured by the parameter $ t_\text{R} $ between sites from the 1st to the 2nd nearest neighbors, where $ \hat{\bm n}_{ij} $ is the unit directional vector for the hopping from site $ j $ to site $ i $; the third term $ H_\text{AFM} $ depicts the antiferromagnetism with sublattice resolution with the vector $ \bm Q $ as either one of the reciprocal lattice vectors; and the last term $ H_\text{SC} $ is the conventional $s$-wave superconducting (SC) pairing with the amplitude $ \Delta $ contributed from the substrate. $ c_i^\dagger = (c_{i\uparrow}^\dagger, c_{i\downarrow}^\dagger) $ is the electron creation operator including spin at site $ i $, and $ \bm{s} = (s_1, s_2, s_3) $ is the spin Pauli matrices.  %and {\color{red} see Sec.~SI of the Supplementary Materials (SM) for the bulk band structure reproduction \footnotemark[1]}.
	%{\color{blue} Junjie, insert the antiferromagnetic time-reversal symmetry analysis and topological invariants.}
	The bulk Hamiltonian \autoref{eq:Hamiltonian} obviously respects the particle-hole (charge conjugation) symmetry $\mathcal{P}$, i.e., $ \mathcal{P} H(\bm k) \mathcal{P}^{-1} = -H(-\bm k) $. %with $ \mathcal{P} = \tau_0 \sigma_1 s_0 K $, where $ K $ is the complex conjugation operation and the three Pauli matrices $\tau$, $\sigma$, and $s$ are responsible for the sublattice, the particle-hole, and the spin degrees of freedom, respectively.
	The AFM order explicitly breaks the time-reversal symmetry $ \Theta $; however, a combined symmetry (or termed as an effective time-reversal symmetry) $ \mathcal{S} = \Theta T_{1/2} $ is preserved, and here $ T_{1/2} = \me^{\mi \bm{k} \cdot (\bm a_1 + \bm a_2)/2} $ denotes the half-translation operator connecting nearest spin-up and spin-down magnetic atoms [see \autoref{fig1:basic_structure_and_behavior}(a)]. This effective time-reversal symmetry $ \mathcal{S} $ is antiunitary with $ \mathcal{S}^2=-\me^{\mi \bm{k} \cdot (\bm a_1 + \bm a_2)} $ and depends on the crystal structure. Moreover, by combining the particle-hole symmetry $ \mathcal{P} $ and effective time-reversal symmetry $ \mathcal{S} $, we can define an AFM chiral symmetry as $ \mathcal{C}=\mathcal{S}\mathcal{P} $, which constrains the bulk Hamiltonian \autoref{eq:Hamiltonian} as $ \mathcal{C} H(\bm k) \mathcal{C}^{-1} = -H(\bm k) $.

    %Specific to the AFM order, the model also enjoys a time-reversal-like symmetry -- hereafter we call it "AFM time-reversal symmetry" -- under the composite operations of time reversal and a half-unit-cell translation: $ \Theta H(\bm k) \Theta^{-1} = H(-\bm k) $, where $ \Theta = \mi \tau_0 \sigma_0 s_2 K T_{1/2}(\bm k) $, where $ T_{1/2}(\bm k) = \me^{\mi\bm{k}\cdot(\bm a_1 + \bm a_2)/2} $. Note that because of this half-unit-cell translation, this AFM time-reversal symmetry relies on the crystallographic structure. Combining them two, one can confirm a chiral symmetry: $ \chi H(\bm k) \chi^{-1} = -H(\bm k) $, where $ \chi = \mathcal{C} \Theta = \mi \tau_0 \sigma_1 s_2 T_{1/2}(\bm k) $. The particle-hole symmetry alone makes the band structures centrosymmetric with respect to $ (E, \bm{k}) = (0, \Gamma) $, while the chiral symmetry makes the band structures with symmetric with respect to $ E = 0 $. As expected, the AFM edge, with the same magnetic configuration as the bulk, hosts zero-energy Majorana flat bands between the projections of the nodal points. On the FM/ZZ edges, however, the AFM time-reversal symmetry is explicitly broken, so is the chiral symmetry. Therefore, the internode edge modes are not protected by $ \chi $ and, therefore, need not be pinned to zero energy. We also note that the ZZ edge is more similar to the AFM structure than the FM, so that the ZZ Majorana edge state has a flatter dispersion than the FM one.}
	
	Since the AFM chiral symmetry $\mathcal{C}$ possesses the crystalline nature, the Majorana edge modes in the AFM topological nodal superconductor are sensitive to edge geometries. When the system terminates with a ribbon geometry along an appropriate crystallographic orientation, as shown in \autoref{fig1:basic_structure_and_behavior}(a), three kinds of edges, i.e., the ferromagnetic (FM), AFM, and zigzag (ZZ) edges, can be formed. The FM and ZZ edges locally break $\mathcal{C}$, which is in contrast to the AFM edge with the $\mathcal{C}$ protection. To reveal topological and transport properties of different edge geometries, we carried out the calculations of local density of states and the Andreev reflection. The values of all parameters in \autoref{eq:Hamiltonian} are chosen to be the same
    %$ (t, t', t'', \mu, t_\text{R}, J, \Delta) = (1.87, 1.53, 1.70, 5.95, 0.85, 5.27, 2.55)~\si{\milli\electronvolt} $
    as in Ref.~\cite{Bazarnik2023NC}, and the technical details are included in the Supplemental Material (SM)~\footnotemark[1]. The topological nodal-point superconducting phase and its Majorana edge modes related to different edges are reproduced as shown in Fig. S1~\footnotemark[1] and \autoref{fig1:basic_structure_and_behavior}(b), respectively.
	%Note that the ZZ edge and the FM edge are mutually perpendicular.
	It is found that these three edges host different Majorana edge modes in terms of degree of dispersiveness and the length spanned in the momentum space. The edge modes on the AFM edge are exactly flat~[see the middle inset of \autoref{fig1:basic_structure_and_behavior}(b)] and pinned to zero energy due to the preserved $ \mathcal{C} $.
	%antiferromagnetic time-reversal symmetry which, in turn, ensures chiral symmetry remains intact.
	However, the projections of bulk nodes on the AFM edge are very close to each other due to the intrinsic structural features, so these flat-band Majorana  edge modes only exist in a rather narrow wave vector window and do not give a ZBCP, as shown in \autoref{fig1:basic_structure_and_behavior}(c). In contrast,
	% to the AFM edge,the FM edge exhibits explicit breaking of the antiferromagnetic time-reversal symmetry.
	the breaking of AFM chiral symmetry $ \mathcal{C} $ leads to the emergence of dispersive Majorana edge modes appearing between the projections of bulk nodes. Remarkably, even with the breaking of $ \mathcal{C} $, the ZZ edge surprisingly hosts Majorana edge modes associated with a nearly flat band and a sizable width of  wave vector window, indicating that a balance between a high density of states and finite group velocities can be achieved. When subjected to a weak coupling to a lead providing incident electrons, the nearly flat-band Majorana edge modes can develop an obvious ZBCP [colored red in \autoref{fig1:basic_structure_and_behavior}(c)].
    %which can develop an obvious ZBCP as displayed in \autoref{fig1:basic_structure_and_behavior}(c)
    % When subjected to a weak coupling to a lead providing incident electrons, these nearly flat-band Majorana edge modes on the ZZ edge
    %And they can develop an obvious ZBCP as displayed in \autoref{fig1:basic_structure_and_behavior}(c). % The ZBCPs result from the combined tunneling process with a net contribution of a normal electron reflection and a local Andreev reflection of a hole, both invoked by an injecting electron from the lead. %See Sec.~SIV of the SM for the transport calculation methods \footnotemark[1].
	
%	%%%%%%%%%%%%%%
%	\footnotetext[1]{See Supplementary Materials at \href{https://journals.aps.org/}{this page} for more details on model construction, pairing correlation for typical minimal models of TSC, and comments on methods for transport calculations, which includes Refs.~\cite{Datta2004,Datta2007,Qiao2007N,Wimmer2009,Reuter2011PRB,Reuter2012CSD,Papior2016,Lima2018PRB}.}
%	%%%%%%%%%%%%%%
	
	%%%%%%%%%%%%%%
	\footnotetext[1]{See Supplementary Materials at \href{https://journals.aps.org/}{this page} for more details on model construction, pairing correlation for typical minimal models of TSC, and comments on methods for transport calculations, which includes Refs.~\cite{Lee1981PRL,Datta2004,Qiao2007N,Reuter2011PRB,Lima2018PRB}.}
	%%%%%%%%%%%%%%
	
	\textit{Pairing Correlation and Spin Polarization of Majorana Edge Modes}.---As depicted above, the dispersion of Majorana edge modes strongly depends on edge geometries in the AFM topological nodal-point superconductor. Specifically, we notice that the nearly flat-band Majorana edge modes on the ZZ edge with breaking the AFM chiral symmetry $\mathcal{C}$ is actually dispersive, which corresponds to finite group velocities though at small magnitudes. Previously related researches all reported that the exactly flat-band Majorana edge modes are dominated by spin-triplet pairing correlations \cite{Yuan2017PRB,he2018magnetic}. Considering that the exactly flat band hosts the entirely zero group velocity but the nearly flat band exhibits a finite group velocity, it is expected that the Majorana edge modes associated with the nearly flat band can give rise to exotic effects. %Previously related researches all reported that Majorana flat-band edge modes are all dominated by spin-triplet pairing correlations \cite{Yuan2017PRB,he2018magnetic}. We notice that, strictly speaking, the ZZ Majorana edge modes under present study are dispersive, which corresponds to finite, though small, group velocities. This condition is entirely missing in the former studies in the similar research context.
    To further reveal this, we examine the pairing correlation of the Majorana edge modes in the present system. In this work, the spin (correlation and polarization) information is encoded in the linear combination coefficients $ d $s of spin Pauli matrices, extracted from the electron-hole block of the Green's function as $ G_\text{eh} = (d_0 s_0 + \bm d \cdot \bm s)(\mi s_2) $~\cite{Gorkov2001PRL,Crepin2015PRB} of a sheet system with a finite width along $ \bm{a}_2 $ but a semi-infinite length along $ \bm{a}_1 $. Then the scalar and vector coefficients $ d_0 $ and $ \bm{d} $ represent, respectively, the amplitudes of spin-singlet and spin-triplet pairing correlations.
	\begin{figure}[htp!]
		\centering
        \includegraphics[width=0.48\textwidth]{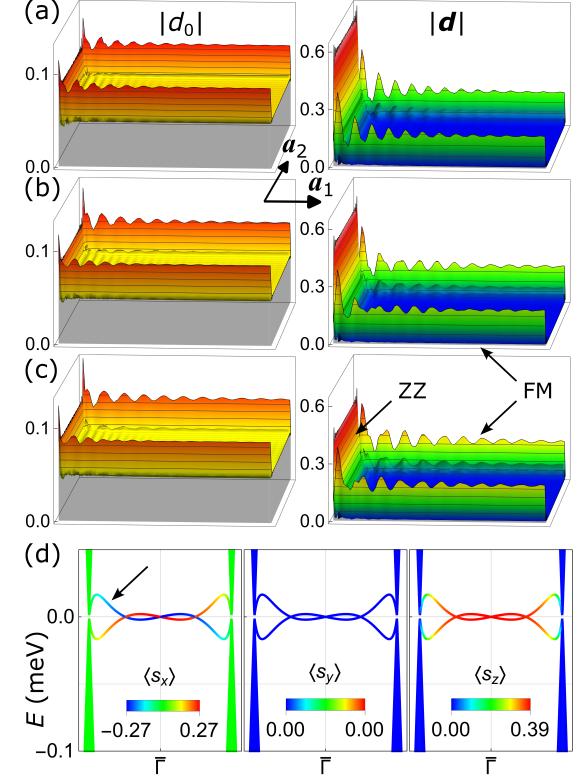}
		\caption{\label{fig2:spin_correlations_and_polarizations} Spin information for the Majorana edge modes. Pairing correlations of the ZZ and the FM Majorana edge modes: (a)-(c) Local spin-singlet $ \abs{d_0} $ (left) and total spin-triplet $ \abs{\bm d} $ (right) correlations for three typical energy values ($ E = 0, 0.03 $, and $ 0.05 $~meV) of the central region; dramatic difference between them does not exist. (d) Components of spin polarization for the pristine ZZ edge modes. % (d) Surface pairing correlation of the ZZ edge mode with resolution of all the four components $ \abs{d_i}\ (i = 0, x, y, \text{ and } z) $ after the attachment of the probe lead.
        }
	\end{figure}
    %And the spin polarization can be obtained by the vector coefficient as $ \bm{s} = \mi(\bm\delta \times \bm\delta^*)/\abs{\bm\delta}^2 $, which has been utilized in \autoref{fig3:spin_AR_and_polarization}(e).
    %How to understand the seemingly counterintuitive results?
    %Now that the ZBCPs seemly behave counterintuitively, it is highly demanding to reveal its mechanism behind.
	% We employ the standard method of Green's function \cite{Gorkov2001PRL,Crepin2015PRB} to extract the pairing correlation of Majorana modes in a sheet system with a finite width but a semi-infinite length. % and the method of scattering matrix \cite{Liu2015PRB} for the open sheet system attached to the (single) lead.
    The calculated results are displayed in \autoref{fig2:spin_correlations_and_polarizations}. Panels (a)-(c) show the local magnitudes of both the spin-singlet ($ \abs{d_0} $) and triplet ($ \abs{\bm d} $) components for three typical energies ($ E = 0 $ meV, 0.03 meV, and 0.05 meV). %($ E = \SIlist{0; .03; .05}{\milli\electronvolt} $), where the edge states dominate.
    One can see that among all the three cases, the spin-singlet component $ \abs{d_0} $ takes finite values all over the central region, but those near the edges are obviously larger than those in the bulk. Meanwhile, the two (ZZ and FM) edges yield mostly the same $ \abs{d_0} $ in magnitudes. On the other hand, the total spin-triplet component $ \abs{\bm d} $ shows a similar trend, except that now the bulk barely contribute and the values on the ZZ edge is much higher than those on the FM edge. More importantly, though smaller, $ \abs{d_0} $ is always of the same order of magnitude as $ \abs{\bm d} $, which is in sharp contrast to the situation of its exactly flat-band counterpart with the negligible spin-singlet pairing amplitude (see Sec.~SII of the SM \footnotemark[1]). Moreover, as the energy increases, the spin-singlet pairing amplitude $ \abs{d_0} $ show slight increase and simultaneously the spin-triplet pairing amplitude $ \abs{\bm d} $ decreases obviously. That is to say, the spin-singlet component can show more prominent manifestations with the increase of energy.
    %the difference between them further becomes smaller. % This message is also reflected in surface version as shown in \autoref{fig2:spin-correlation}(b), where the spin-singlet correlation indeed exeeds each individual component of spin-triplet one in most part of the energy window, even with the probe lead attached, although the three spin-triplet components do not behave the same.
    Additionally, as shown in \autoref{fig2:spin_correlations_and_polarizations}(d), the left-upper branch of ZZ edge mode (indicated by the black arrow) bears a spin polarization of finite negative and positive components, respectively, in the $ x $- and $ z $-direction; but a vanishing one for $ s_y $. Namely, around the zero energy this ZZ edge mode has a spin polarization with an acute polar angle and an azimuthal angle of the largest magnitude. Specifically, when the self energy modification of a probe lead is included, this ZZ edge mode would possess a spin polarization with the orientation $ (\theta_\text{s}, \phi_\text{s}) \approx (\pi/3, \pm\pi) $ (see the details in Sec.~SIII of the SM \footnotemark[1]).
	
	%Though surprising, the status quo is not beyond comprehension. It has been established in Ref.~\cite{Sato2009PRL} that a topologically nontrivial superconductivity with gapless dispersive chiral Majorana edge modes originate from a spin-singlet pairing correlation and owns considerable components of (effective) spin-triplet pairing correlation seen in a proper representation. This is supported by later works like Refs. \cite{Liu2015PRB,Yuan2017PRB,Zhang2017PRB}. However, the aforementioned effective spin-triplet components are proportional to the strength of the (Rashba) SOC effect and all of them deal with zero-energy Majorana boundary (end or edge) states without a transverse (perpendicular to the direction of the lead) velocity. This in fact leaves room, though not noticed before, for different behaviors for dispersive Majorana edge modes like in this work.
	It is evident that the spin-singlet component of pairing correlation remains largely unsuppressed for the nearly flat-band Majorana edge modes on the ZZ edge. This contrasts significantly with previously studied systems with exactly flat-band Majorana edge modes, where the spin-singlet contribution is negligible compared with the spin-triplet component \cite{Yuan2017PRB,he2018magnetic}. The key to understanding this discrepancy lies in the AFM chiral symmetry of the system~\cite{Bazarnik2023NC,Kieu2023PRB}. While the 2D bulk model respects this symmetry, it is spatially dependent, meaning that introducing edges can disrupt it. The AFM chiral symmetry is conserved by the AFM edges, but is broken by the the ZZ edges. The same information is embodied in Majorana edges modes that those for the AFM edge are flat but those for the latter become dispersive [\autoref{fig1:basic_structure_and_behavior}(b)]. To verify this issue, we also change the parameters in \autoref{eq:Hamiltonian} to enlarge the wave vector window of flat-band Majorana edge modes on the AFM edge. As expected, the results show that the spin-singlet pairing amplitude is also negligible (see Fig. S2 in the SM \footnotemark[1]).
%Given the narrow wave vector window of the flat-band Majorana edge modes in our system, as previously noted, we are not computing the pairing correlation on the AFM edge.
	
\begin{figure}[htp!]
		\centering
		\includegraphics[width=0.48\textwidth]{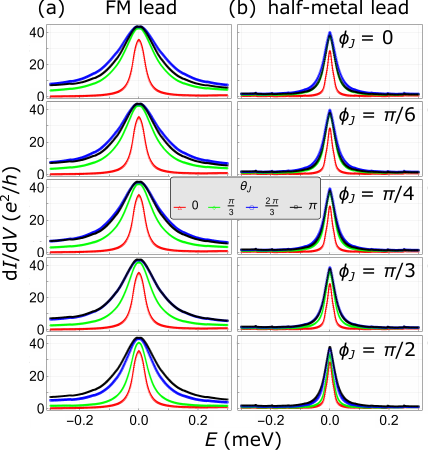}
		\caption{\label{fig3:spin_AR_ZZ} Spin-dependent ZBCPs from the ZZ Majorana edge modes for (a) FM and (b) Half-metal probe lead. $ (\theta_J, \phi_J) $ specify the orientation of the spin polarization of the incident electron.
        % (c) The expectation values for the $ s_z $ in a ZZ-edged nanoribbon structure, corresponding to the central region of the ZZ spectral function in \autoref{fig1:basic_structure_and_behavior}(b); so the edge mode [from left upper to right lower, see \autoref{fig1:basic_structure_and_behavior}(b) ZZ] has a spin with a majority of positive $ z $-component.
        }
	\end{figure}

	\textit{Manifestation in Spin-Selective Andreev Reflection}.---In the following, we show that the nearly flat-band Majorana edge modes have unique Andreev reflection features when the incident electrons are spin polarized, facilitating its experimental detection. To start with, we first let the electrons from a FM probe lead inject onto the ZZ edge of a central rectangular region,
    % in a geometry with a finite width but a semi-infinite length,
    because the Majorana edge modes on the ZZ edge contribute to the most prominent ZBCP as shown in \autoref{fig1:basic_structure_and_behavior}(c). The polarization of the incident electrons are characterized by the polar angle $ \theta_J $ and the azimuthal angle $ \phi_J $, defined with respect to the directions of $ (\bm a_1 \times \bm a_2) $ and $ \bm a_1 $, respectively. The corresponding results are shown in \autoref{fig3:spin_AR_ZZ}(a). It is found that for a completely out-of-plane polarized spin, i.e., $ \theta_J = 0 $ (colored in red) or $ \theta_J = \pi $ (colored in black),
    %We note that for a completely out-of-plane polarized spin ($ \theta_J = 0 \text{ or } \pi $),
    the change of $ \phi_J $ makes no difference. %As a result, the curves with red triangles are totally the same, so are the one with black squares.
    And they can serve as the eye-guiders.
    Starting from the $ \phi_J = 0 $ case, one can see that the curve of $\theta_J = 2\pi/3 $ (colored in blue) is above the one of $\theta_J = \pi $ (colored in black) and the one of $\theta_J = \pi/3 $ (colored in green) lies below. %and the one with green diamonds below.
    The increase of the azimuthal angle pushes the blue and the green curves close to the lowest red one. The difference is that although both of them are decreasing overall, the height of the blue peak largely maintains, but the green peak reduces. More importantly, no matter how the azimuthal angle changes ($ \phi_J \in [0, \pi/2] $), the highest and widest ZBCP always comes from some obtuse polar angles ($ \theta_J = 2\pi/3, \pi $).
    %On the other hand, the spin polarization of the edge modes is shown in \autoref{fig3:spin_AR_ZZ}(c), displaying a major positive out-of-plane component together with a small negative $ x $-component, which would increase with the attachment of the probe lead to the edge, see Sec.~SIII of the SM for the spin polarization data obtained with the probe lead attached \footnotemark[1].
    Combining all the above messages with the spin polarization displayed in the previous section [see~\autoref{fig2:spin_correlations_and_polarizations}(d)], we find that when the spin polarization of the injecting electron in the probe lead is (approximately) opposite to that of the Majorana edge mode in the central scattering region, the heights and widths of the ZBCPs are maximized. Moreover, consistent behaviors from a half-metal probe lead are shown in \autoref{fig3:spin_AR_ZZ}(b), where the injection electrons are entirely spin polarized \cite{Keizer2006N} (see Sec.~SIII of the SM for adding magnetism into the probe lead \footnotemark[1]). Now although the maximal heights of the ZBCPs are slightly smaller, the peaks are overall thinner in width, making them display more visual prominence. To summarize, our findings demonstrate that while the nearly flat-band Majorana edge modes on the ZZ edge lead to a wide ZBCP, the coexistence of spin-singlet and triplet components of the pairing correlation results in a spin-dependent Andreev reflection that deviates from that induced by the exactly flat-band Majorana edge modes protected by the AFM chiral symmetry  $\mathcal{C}$.

	\textit{Summary and Discussion}.---%We have investigated the the spin features of Majorana edge modes in a topological nodal-point superconductor with antiferromagnetism, in which the spectrum of Majorana edge modes strongly depends on the edge geometries with specific symmetry broken. %analyzed the spin features of Majorana edge modes in a realistic 2D topological nodal-point superconductor with antiferromagnetism.
	We investigated the spin characteristics of Majorana edge modes in a topological nodal-point superconductor with antiferromagnetism, where edge geometry and symmetry breaking significantly influence the spectrum features of Majorana edge modes. We discovered that breaking the antiferromagnetic chiral symmetry leads to non-negligible spin-singlet pairing correlations in nearly flat-band Majorana edge modes. This challenges the previous assumption that exactly flat-band Majorana edge modes in 2D topological nodal superconductors are solely dominated by spin-triplet pairing. Our finding hinges on the ability of nearly flat-band Majorana edge modes to balance high density of states with finite group velocities. Furthermore, we demonstrated unique spin-selective Andreev reflection behaviors in these Majorana edge modes.
	
	Our results emphasize the importance of considering the potential manifestation of spin-singlet pairing correlations in transport processes like Andreev reflection, especially when the dispersion of Majorana edge states connecting bulk node projections deviates from strict flatness. This work contributes to a deeper understanding of the spin-dependent mechanisms governing Majorana fermions.
	
	\textit{Acknowledgement}.---This work was supported by the National Natural Science Foundation of China (NSFC, Grants No. 12222402, No. 92365101, No. 12074108, and 12347101), the Chongqing Natural Science Foundation (Grants No. CSTB2023NSCQ-JQX0024 and No. CSTB2022NSCQ-MSX0568).

	% \bibliography{../../../Library/afm_tnpsc}
	%\bibliography{afm_tnpsc}
	\bibliographystyle{apsrev4-1}

\end{document}